\documentclass[superscriptaddress,amsmath,amssymb,aps,prl,twocolumn]{revtex4-1}

\usepackage{graphicx}
\usepackage{dcolumn}
\usepackage{xcolor}
\usepackage{mathrsfs}
\usepackage{xspace}
\usepackage{upgreek}

\newcommand{\ket}[1]{{\left| {#1} \right\rangle}}
\newcommand{\bra}[1]{{\left\langle {#1} \right|}}

\newcommand{\WignerThreej}[6]{{\left( \begin{array}{ccc} {#1} & {#2} & {#3}  \\  {#4} & {#5} & {#6} \end{array} \right)}}

\newcommand{\boldgreek}[1]{{\mbox{\boldmath$ {#1} $}}}
\newcommand{\boldgreeksmall}[1]{{\mbox{\scriptsize\boldmath$ {#1} $\normalsize}}}
\newcommand{\shrinkify}[1]{\textstyle {#1} \displaystyle}

\newcommand{\SigmaSigma}{\ensuremath{{}^1\Sigma^+ \leftarrow {}^1\Sigma^+}\xspace}
\newcommand{\SigmaPi}{\ensuremath{{}^1\Sigma^+ \leftarrow {}^1\Pi}\xspace}

\DeclareMathSymbol{\shortminus}{\mathbin}{AMSa}{"39}

\newcommand{\YFplus}{YF$^+$\xspace}
\newcommand{\pgopher}{\textsc{pgopher}\xspace}
\newcommand{\Tkin}{E_\mathrm{kin}}

\begin{document}

\title{Photon spin molasses for laser cooling molecular rotation}

\author{W. C. Campbell}
\affiliation{Department of Physics and Astronomy, University of California Los Angeles, Los Angeles, California 90095, USA}
\affiliation{UCLA Center for Quantum Science and Engineering, University of California, Los Angeles, Los Angeles, California 90095, USA}
\affiliation{Challenge Institute for Quantum Computation, University of California, Los Angeles, Los Angeles, California 90095, USA}
\author{B. L. Augenbraun}
\affiliation{Department of Physics, Harvard University, Cambridge, Massachusetts 02138, USA}
\affiliation{Harvard-MIT Center for Ultracold Atoms, Cambridge, Massachusetts 02138, USA}

\date{November 6, 2021}

\begin{abstract}
Laser cooling of translational motion of small molecules is performed by addressing transitions that ensure spontaneous emission cannot cause net rotational excitation.  This will not be possible once the rotational splitting becomes comparable to the operational excitation linewidth, as will occur for large molecules or wide bandwidth lasers.  We show theoretically that in this regime, angular momentum transfer from red-detuned Doppler cooling light can also exert a damping torque on linear molecules, cooling rotation to the same Doppler limit (typically $\approx \! 500 \mbox{ }\upmu \mathrm{K}$ for molecules with $\approx \! 10$~ns excited-state lifetimes).  This cooling process is derived from photon spin, and indicates that standard optical molasses can also cool molecular rotation with no additional experimental resources.

\end{abstract}

\maketitle

The mechanical control of gas-phase atoms and molecules by light has its origins in two classic experiments from the 1930s.  First, in 1933, Frisch demonstrated deflection of atoms by the transfer of linear momentum from light \cite{Frisch1933Experimenteller}.  Decades later, following the development of tunable lasers that could address atomic resonances with high spectral purity, this was adapted to perform laser cooling \cite{Haensch1975Cooling,Wineland1975Proposed} and has been employed to cool the translational motion of gas phase atoms \cite{Wineland1978RadiationPressure,Phillips1982Laser,Chu1985ThreeDimensional} and molecules \cite{Shuman2009DiatomicRadiativeForce,Kozyryev2017Sisyphus,Mitra2020Direct} to sub-milikelvin temperatures.  The cooling force in these cases is derived from the translational momentum of the photon.  For Doppler cooling, the minimum attainable temperature is set by a balance between the damping force and the stochastic heating caused by fluctuations of the direction of the momentum change from scattering, typically $\approx 500 \text{ }\upmu \mathrm{K}$ for transitions with $\sim$10~ns lifetimes.

While translational laser cooling has had a transformative effect on the possibilities for precision measurements and quantum mechanical applications with atoms, cooling of the angular degrees of freedom in atoms (\emph{viz.}, the angular momentum $J$ and its laboratory-frame projection $M$) garners less attention, likely because it is fairly easy to optically pump atomic population into single quantum states. This is not an easy task, however, for molecules---particularly for large molecules. The thermal distribution of population among molecular rotational levels can be a substantial reservoir of entropy, and line broadening from thermal fluctuations and finite spontaneous emission lifetime begins to obscure quantized rotation when the excitation linewidth, $\gamma_\mathrm{eff}$, exceeds the spectral shift from rotation, $\gamma_\mathrm{eff} > 4 \pi B \langle J \rangle$, where $B$ is the molecular rotational constant in units of cyclic frequency.

To control molecular rotation with light in this rotationally unresolved regime, we can look not to the linear momentum transfer observed by Frisch, but to a demonstration made three years later of a different mechanical effect exerted by photons.  In a 1936 experiment \cite{Beth1936Mechanical}, Beth demonstrated the \emph{torque} exerted by transfer of light's intrinsic \emph{angular} momentum, which we now identify as photon spin. Following the recent advances in molecular laser cooling, it appears natural to ask: what sort of \textit{rotational} cooling can be achieved if one follows Beth's experiment to seek an analog to translational Doppler cooling?

Here, we show theoretically that red-detuned light can exert a damping torque on rotating linear molecules, even in the rotationally unresolved regime, owing to the preferential absorption of photons with spin projection anti-aligned to the molecular angular momentum.  We present this effect using a general framework that can be applied to either translational or rotational cooling of classical degrees of freedom. This treatment stresses the connections between the two types of cooling and explains why both share the same Doppler limit.  We then extend this to some quantized linear molecules and present the results of numerical simulations that illustrates cooling of quantum mechanical rotation.  This work complements recent applications of related ideas for rotational cooling of classical nanoparticles \cite{Battacharya2007Using,RomeroIsart2010Toward,Shi2016Optomechanics,Hoang2016Torsional,Pan2021Rotational}.  In all cases, we limit the scope of this analysis to rotation by assuming that vibrational branching can be managed similarly to how it has been for some smaller molecules \cite{Kozyryev2017Sisyphus,Mitra2020Direct}.  These results suggest that optical molasses may be capable of cooling the rotation of large molecules to sub-millikelvin temperatures for applications including spectroscopy, precision measurement, and ultracold chemistry.

The rest of this paper is organized as follows. We begin with a general framework for modelling the mechanical effect of laser light on gas-phase absorbers.
This model is used to show that the Doppler cooling limit is the same for rotation as it is for translation, but we find that the cooling rate can be orders of magnitude higher for rotation for sub-wavelength molecules.
We then describe how photon spin molasses works for a few types of linear, quantized molecules, and present the results of numerical simulations that include complexity found in real molecules.
We find that in some cases, the addition of realistic molecular features leads to rotational population gathering around finite $J$ rather than near $J=0$, but orders-of-magnitude increases in peak phase-space density demonstrate that cooling still occurs.

\section{Doppler cooling of classical degrees of freedom with narrow linewidth laser light}
We begin by considering a molecule (or atom) with a continuous vector momentum $\boldgreek{\vec{\pi}}$ associated with a degree of freedom to be cooled. For translation, this is the translational momentum $\boldgreek{\vec{\pi}} = \vec{\mathbf{p}}$ and for rotation it is the angular momentum $\boldgreek{\vec{\pi}} = \vec{\mathbf{L}}$. Further correspondences are listed in Table \ref{tab:correspondence}. The momentum is related to the kinetic energy by the degree of freedom's inertia $\mu$ through $\Tkin = |\boldgreek{\vec{\pi}}|^2/(2\mu)$.  In the interest of simplicity, for the rotational case, we will assume a spherical top in this section so that $\mu$ is isotropic in both cases.

\begin{table}
    \centering
    \begin{tabular}{r |c|c|c}
    \hline \hline
         quantity & general & translation & rotation \\
        \hline
         momentum & $\boldgreek{\vec{\pi}}$ & $\vec{\mathbf{p}}$ & $\vec{\mathbf{L}}$ \\
         inertia & $\mu$ & $m$ & $I$ \\
         velocity & $\boldgreek{\vec{\pi}}/\mu$ & $\vec{\mathbf{v}}$ & $\boldgreek{\vec{\omega}}$ \\
        photon momentum & $\hbar \boldgreek{\vec{\kappa}}$ & $\hbar \vec{\mathbf{k}}$ & $\hbar \mathbf{\hat{k}}$ \\
        recoil frequency & $\omega_\mathrm{r}$ & $\hbar k^2/(2m)$ & $\hbar/(2I)$ \\
        capture speed & $|\boldgreek{\vec{\pi}}_\mathrm{c}|/\mu$ & $\gamma/k$ & $\gamma$ \\
        mechanical damping & $\mathrm{d} \boldgreek{\vec{\pi}}/\mathrm{d} t$ & force & torque\\
          \hline \hline
    \end{tabular}
    \caption{\label{tab:correspondence}Correspondence table for translational and rotational Doppler cooling.}
\end{table}

For photon scattering, the molecule will have an atom-like optical transition with a rest-frame resonant frequency $\omega_\mathrm{mol}$ and spectral width $\gamma$. For a molecule with initial momentum $\boldgreek{\vec{\pi}}_0$, upon absorbing a laser photon with momentum $\hbar \boldgreek{\vec{\kappa}}$, its kinetic energy is changed by an amount
\begin{equation}
    \Delta \Tkin = \hbar \omega_\mathrm{r} + \hbar \frac{\kappa \, |\boldgreek{\vec{\pi}}_0|}{\mu} \cos\left( \theta \right) \label{eq:DeltaT}
\end{equation}
where $\theta$ is the angle between $\boldgreek{\vec{\pi}}_0$ and $\boldgreek{\vec{\kappa}}$ and the recoil frequency is given by $\omega_\mathrm{r} \equiv \hbar \kappa^2/(2\mu)$.

The first term in Eq.~(\ref{eq:DeltaT}), which is always positive, is the recoil energy.  The second term can be interpreted as the energy change from the apparent shift of the temporal frequency of the light in the body-fixed frame of the molecule, which is to say a Doppler-type shift.  In order for the applied light to be perfectly resonant with the molecular transition, the laser must be detuned from $\omega_\mathrm{mol}$ to account for the shift from Eq.~(\ref{eq:DeltaT}).

We can conclude two things immediately. First, scattering will only be significant for molecules with Doppler-shifted resonances within about one linewidth of the applied light's frequency, which allows us to define the capture momentum, $|\boldgreek{\vec{\pi}}_\mathrm{c}| \equiv \mu \gamma/\kappa$. Second, we find that the transitions that damp the molecule's momentum ($\boldgreek{\vec{\kappa}}\cdot \boldgreek{\vec{\pi}}_0 < 0$) have lower resonance frequencies than those that amplify the momentum.  Neglecting the small contribution from the recoil shift, this implies that red-deutned light will preferentially damp motion upon absorption.  Subsequent spontaneous emission for this classical moleucule will be equally likely to damp as amplify, and will do neither on average.

We will model the scattering of laser light as a separate process for each distinct applied value of $\boldgreek{\vec{\kappa}}$ that can be approximated by the steady-state two level system scattering rate
\begin{equation}
    \Gamma(\boldgreek{\vec{\kappa}}) = \frac{\frac{\gamma}{2}s_\boldgreeksmall{\vec{\kappa}} }{1 + s_\mathrm{tot} + \frac{4}{\gamma^2}\left(\delta' - \omega_\mathrm{r} - \boldgreek{\vec{\kappa}}\cdot \boldgreek{\vec{\pi}}/\mu \right)^2} \label{eq:SteadyStateScatteringRate}
\end{equation}
where $s_\boldgreeksmall{\vec{\kappa}}$ is the resonant saturation parameter of that particular component of the applied light, $s_\mathrm{tot}$ is the sum of resonant saturation parameters from all applied light, and $\delta' \equiv \omega_\mathrm{laser} - \omega_\mathrm{mol}$ is the lab-frame laser detuning (assumed to be the same for all of the light).  From here on we will absorb the recoil frequency into the definition $\delta \equiv \delta' - \omega_\mathrm{r}$.

\begin{figure}
    \centering
    \includegraphics[width=0.45\textwidth]{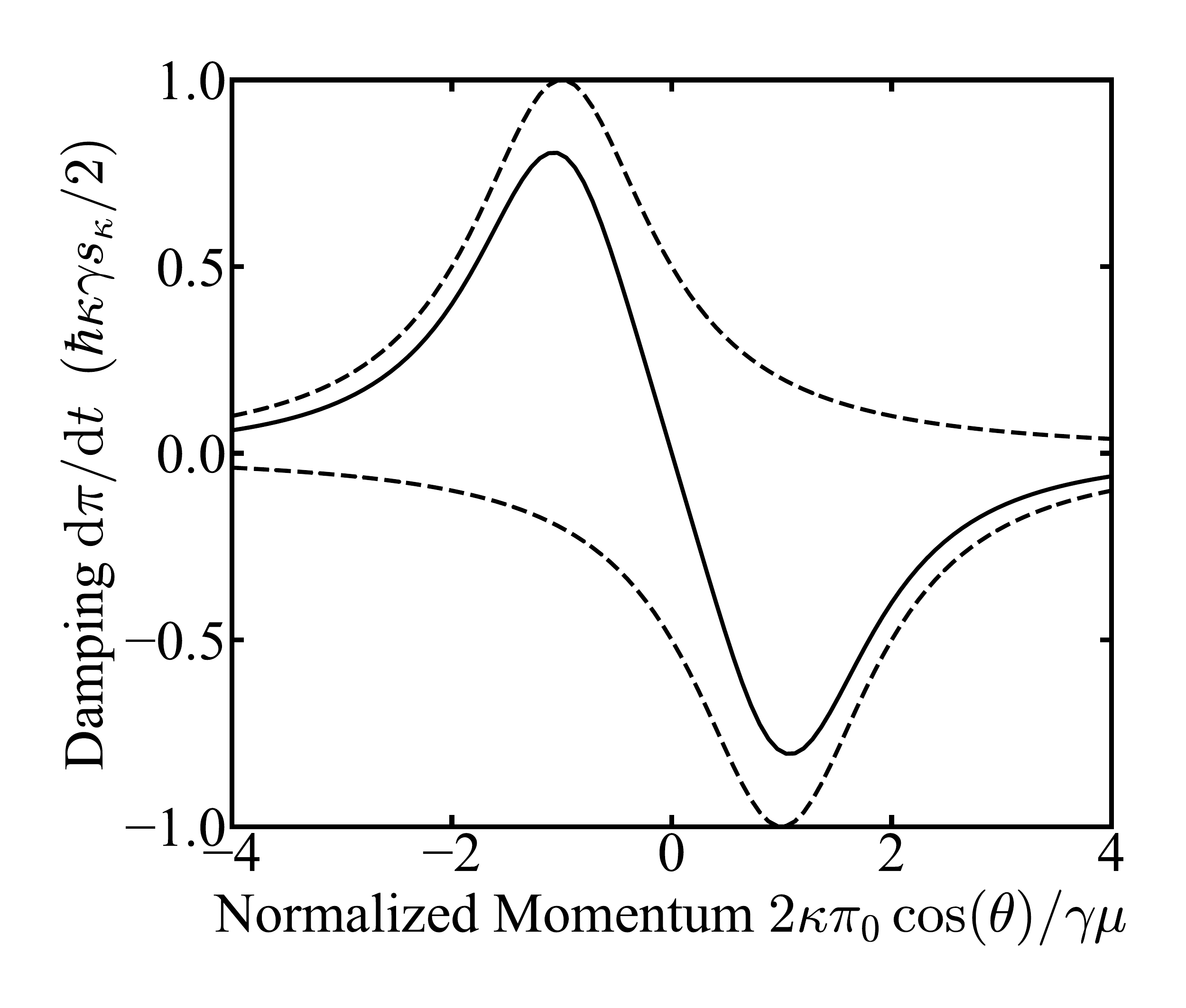}
    \caption{Mechanical damping from Doppler cooling of a lifetime-broadened absorber in 1 dimension. The solid curve shows the net damping from Eq.~(\ref{eq:MechanicalDampingExact}) with detuning $\delta = -\gamma/2$ in the low-intensity limit (and neglecting the recoil shift).  Dashed curves show the contribution from the individual $+\boldgreek{\vec{\kappa}}$ and $-\boldgreek{\vec{\kappa}}$ components.}
    \label{fig:SimpleDamping}
\end{figure}

If only two beam components are applied with anti-parallel $\boldgreek{\vec{\kappa}}$-vectors and identical intensity ($s_{\boldgreeksmall{\vec{\kappa}}_i} = s_0$), the mechanical effect on the molecules can be written as
\begin{align}
    \frac{\mathrm{d}\boldgreek{\vec{\pi}}}{\mathrm{d}t} &= \hbar \boldgreek{\vec{\kappa}} \Big( \Gamma(+\boldgreek{\vec{\kappa}}) - \Gamma(-\boldgreek{\vec{\kappa}})  \Big) \label{eq:MechanicalDampingExact}\\
    & \approx  \hbar \boldgreek{\vec{\kappa}} \frac{8 \gamma^3 \delta s_0}{\left[\gamma^2(1 + 2 s_0) + 4 \delta^2  \right]^2} \frac{\kappa \, |\boldgreek{\vec{\pi}}|}{\mu} \cos\left( \theta \right) \label{eq:MechanicalDampingApprox}
\end{align}
where (\ref{eq:MechanicalDampingApprox}) applies in the low-temperature limit ($|\Delta \Tkin| \ll \hbar |\delta|,\hbar \gamma$).  Figure \ref{fig:SimpleDamping} shows the damping from Doppler cooling given by Eq.~(\ref{eq:MechanicalDampingExact}) in dimensionless units with the laser detuned by half a linewidth red of rest-frame resonance, reproducing the the shape that is familiar from translational optical molasses \cite{Lett1989Optical}.

For low temperatures, if six equally intense beam components are applied with $\boldgreek{\vec{\kappa}}$-vectors aligned and anti-aligned with each of the Cartesian directions and we can neglect optical interference between different components, the damping is aligned with the vector momentum,
\begin{equation}
    \frac{\mathrm{d}\boldgreek{\vec{\pi}}}{\mathrm{d}t} = -\alpha \frac{\boldgreek{\vec{\pi}}}{\mu}, \label{eq:dpidt}
\end{equation}
where the damping coefficient $\alpha$ is given by
\begin{equation}
    \alpha = -\delta \frac{8 \hbar \gamma^3 s_0 \kappa^2}{\left[ \gamma^2 (1 + 6s_0) + 4 \delta^2 \right]^2}.
\end{equation}
Equation (\ref{eq:dpidt}) allows us to identify the energy damping time constant, $\tau_E = \mu/(2\alpha)$, which sets the timescale for the momentum to approach equilibrium.  If we compare rotational and translational damping of a molecule with characteristic size $\ell$, the rotational damping rate is faster by a factor of $\approx (\lambda/\ell)^2$, which indicates that in the cold regime, rotational energy will be cooled faster than translational energy for sub-wavelength molecules.

To get an expression for the Doppler limit, we can write the low temperature cooling power, which is negative for cooling and positive for amplification, as
\begin{equation}
    \dot{E}_\mathrm{cool} = \frac{\mathrm{d}\boldgreek{\vec{\pi}}}{\mathrm{d}t} \cdot \frac{\boldgreek{\vec{\pi}}}{\mu} =
    -\tau_E^{-1} \frac{|\boldgreek{\vec{\pi}}|^2}{2\mu}.
\end{equation}

At steady state (in the absence of other sources of heating), this cooling power will be balanced by a heating power that is due to the randomness in the momentum transfer upon absorption and emission.  If we consider for a moment the case of rotational cooling with a single, linearly-polarized beam, each incident photon is in a superposition of $\boldgreek{\vec{\kappa}} = + \mathbf{\hat{k}}$ and $\boldgreek{\vec{\kappa}} = - \mathbf{\hat{k}}$ spin states.  If we were to model each absorption event as a quantum measurement of the photon spin projection along the $\mathbf{\hat{k}}$ direction, we would identify this fluctuation as quantum projection noise.  The same is true for translational cooling by long-coherence-length photons in superpositions of $\pm \vec{\mathbf{k}}$, though the connection to projection noise is perhaps less obvious.  Nonetheless, this illustrates that the heating can be thought of conceptually as coupling to quantum fluctuations.

As the molecules approach the Doppler limit, the absorption (and emission) of each allowed value of $\boldgreek{\vec{\kappa}}$ becomes equally likely, and we can write the heating rate as
\begin{equation}
    \dot{E}_\mathrm{heat} = \frac{1}{2\mu} \frac{\mathrm{d}}{\mathrm{d}t} \langle \boldgreek{\vec{\pi}}^2 \rangle = \frac{\mathcal{D}}{\mu}
\end{equation}
where the momentum diffusion constant is given by
\begin{equation}
    \mathcal{D} = \frac{3 \hbar^2 \gamma^3 \kappa^2 s_0}{\gamma^2 (1 + 6 s_0) + 4 \delta^2}.
\end{equation}
Equilibrium will be reached when $\dot{E}_\mathrm{cool} + \dot{E}_\mathrm{heat}=0$.  For this isotropic three-dimensional model, the total kinetic energy will be $\frac{3}{2}k_\mathrm{B}T_\mathrm{D}$, which yields the well known Doppler cooling limit \cite{Phillips1998Nobel,Leibfried2003Quantum},
\begin{equation}
    T_\mathrm{D} = -\frac{\hbar \gamma}{4 k_\mathrm{B}} \left( (1 + 6s_0)\frac{\gamma}{2 \delta} + \frac{2 \delta}{\gamma} \right).\label{eq:DopplerLimit}
\end{equation}
Since the Doppler limit from this classical model does not contain any parameters that depend upon the nature of the degree of freedom being cooled (\textit{i.e.}~translational \textit{vs.}~rotational), the familiar sub-millikelvin scale for translational Doppler cooling should also apply to the achievable rotational temperature.

\begin{figure}
    \centering
    \includegraphics[width=0.45\textwidth]{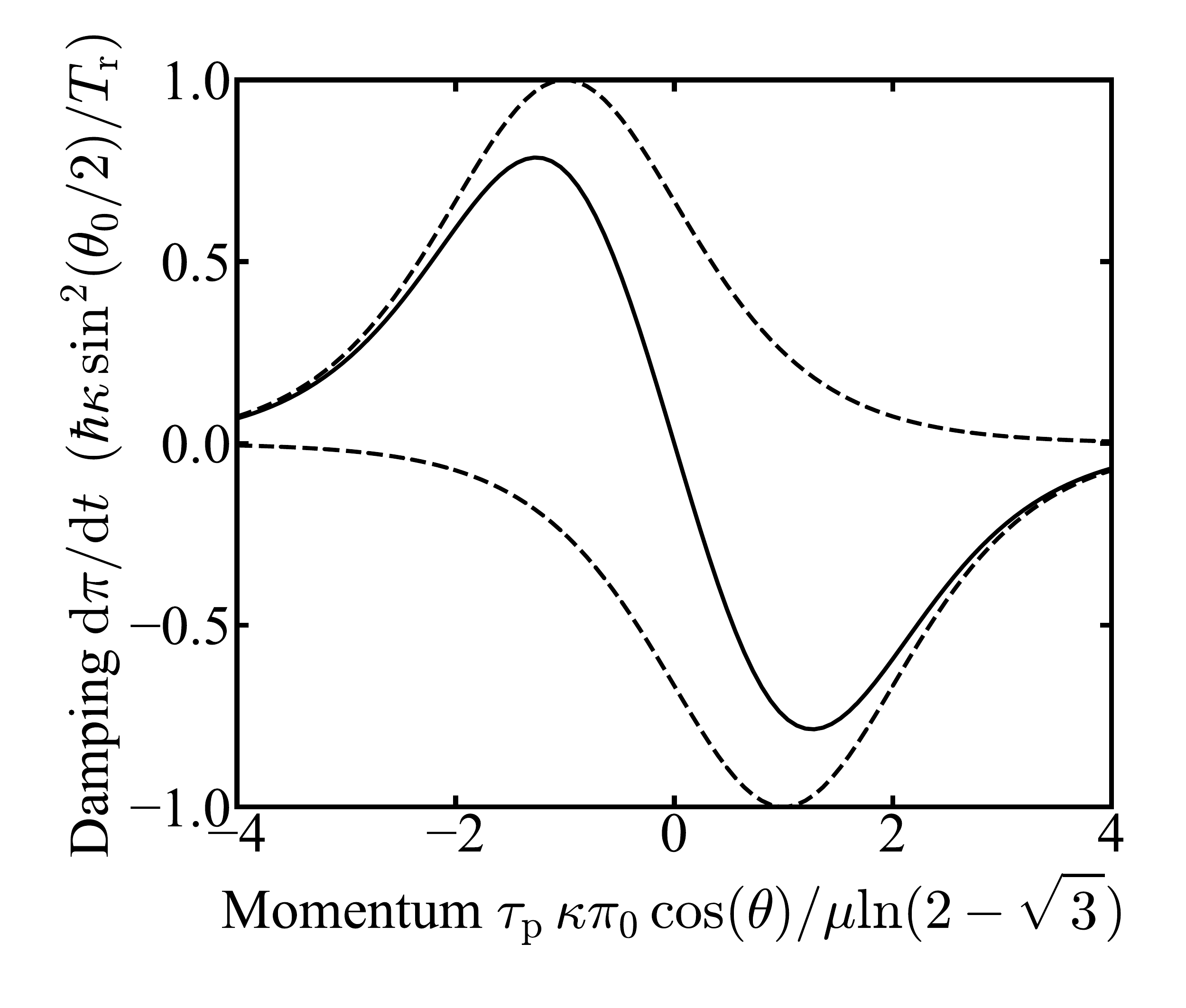}
    \caption{Mechanical damping from Doppler cooling by a hyperbolic secant spectrum in 1 dimension. The solid curve shows the net damping with detuning $\delta = \frac{1}{\tau_\mathrm{p}}\ln(2 - \sqrt{3})$ in the low-intensity limit (and neglecting the recoil shift).  Dashed curves show the contribution from the individual $+\boldgreek{\vec{\kappa}}$ and $-\boldgreek{\vec{\kappa}}$ components.}
    \label{fig:SimpleSechDamping}
\end{figure}

\subsection{Cooling with wide bandwidth light}
The second context in which the rotationally unresolved regime can be found is when the excitation linewitdth $\gamma_\mathrm{eff}$ is set by the linewidth of the laser $\Delta \omega_L$, as opposed to the natural lifetime of the optical transition $\gamma$.  For the rotational cooling of moderate-sized molecules with $B$ on the order of $\mathrm{GHz}$, a picosecond mode-locked laser could provide light capable of addressing many rotational states at once.

Here, we consider the case where the laser applies a train of pulses that are separated by a repetition period $T_\mathrm{r}$ that is much longer than $1/\gamma$ to avoid comb tooth effects that can trap population at high-lying fixed points (\textit{viz.}\ $\mathrm{sech}(\gamma T_\mathrm{r}/2) \ll 1$) \cite{Ip2018Phonon}.  Since inter-pulse coherence can be ignored in this regime, we model the spectrum as the single-pulse spectrum (as opposed to an optical frequency comb) from transform-limited pulses of the form $\mathbf{E}(t) \propto \mathrm{sech}(\pi t/\tau_\mathrm{p})$, where $\tau_\mathrm{p} \ll T_\mathrm{r}$ is the pulse duration.

For a pulse whose center frequency is detuned from resonance by $\delta$, the Hamiltonian for a two-level absorber is given by
\begin{equation}
    H(t) = -\frac{\delta}{2} \sigma_Z - \frac{\Omega_0}{2}\, \mathrm{sech}\! \left( \frac{\pi t}{\tau_\mathrm{p}}\right) \sigma_X, \label{eq:SechHamiltonian}
\end{equation}
where $\Omega_0$ is the peak Rabi frequency of the interaction and the $\sigma_i$ are the Pauli operators for the two-level system.  The excitation probability for a single pulse illuminating a ground-state absorber of this form is given by the Rosen-Zener solution \cite{Rosen1932Double},
\begin{equation}
    P_\mathrm{ex} = \sin^2\!  \left(\frac{\theta_0}{2} \right)\mathrm{sech}^2 \! \left( \frac{\delta \,\tau_\mathrm{p}}{2} \right) \label{RosenZener}
\end{equation}
where the pulse area is $\theta_0 \equiv \int \! \mathrm{d}t' \,\Omega_0 \, \mathrm{sech}(\pi t'/\tau_\mathrm{p}) = \Omega_0 \tau_\mathrm{p}$.  For a system with $N$ excited states within the excitation bandwidth, we can model each using Eq.~(\ref{RosenZener}) so long as $N \theta_0 \ll \pi$.

For Doppler cooling, we can apply the same formalism as before if we replace the lifetime-broadened steady state scattering rate, Eq.~(\ref{eq:SteadyStateScatteringRate}), with
\begin{equation}
    \Gamma(\boldgreek{\vec{\kappa}}) = \frac{ \sin^2 \! \left( \shrinkify{\frac{\theta_0}{2}}\right)}{T_\mathrm{r}} \mathrm{sech}^2 \! \left( \frac{\delta \,\tau_\mathrm{p}}{2} - \frac{\tau_\mathrm{p}}{2 \mu} \boldgreek{\vec{\kappa}} \cdot \boldgreek{\vec{\pi}}\right).
\end{equation}
The damping effect of this scattering force for 1D molasses (see Eq.~(\ref{eq:MechanicalDampingExact})) is shown in Fig.~\ref{fig:SimpleSechDamping} for a detuning $\delta = \frac{1}{\tau_\mathrm{p}} \ln(2 - \sqrt{3})$.  The shape of the response mimics many of the aspects of the lifetime-limited damping shown in Fig.~\ref{fig:SimpleDamping}, and many of the conclusions of the previous section will be qualitatively consistent with the replacement $\gamma \rightarrow 1/\tau_\mathrm{p}$ (such as the capture momentum, which in this case would be $|\boldgreek{\vec{\pi}}_\mathrm{c}| \equiv \mu/(\kappa \tau_\mathrm{p})$).

In the cold regime ($|\boldgreek{\vec{\kappa}} \cdot \boldgreek{\vec{\pi}}| \ll \delta $) with six optical fields to provide $\boldgreek{\vec{\kappa}}$ both parallel and antiparallel to each Cartesian direction in space, the damping effect of the light is isotropic and yields the energy damping rate
\begin{equation}
    \frac{1}{\tau_E} = \frac{2 \alpha}{\mu} = -\frac{4 \hbar \kappa^2 \tau_\mathrm{p} \sin^2 \! \left( \shrinkify{\frac{\theta_0}{2}}\right)}{\mu T_\mathrm{r}}\, \mathrm{tanh}\! \left( \frac{\delta \,\tau_\mathrm{p}}{2}\right) \mathrm{sech}^2 \! \left(  \frac{\delta \, \tau_\mathrm{p}}{2}\right) \! .\label{eq:SechCoolingRate}
\end{equation}
The stochastic heating power is given by
\begin{equation}
    \dot{E}_\mathrm{heat} = \frac{6 \hbar^2 \kappa^2  \sin^2 \! \left( \shrinkify{\frac{\theta_0}{2}}\right)}{\mu T_\mathrm{r}} \mathrm{sech}^2\! \left( \frac{\delta \, \tau_\mathrm{p}}{2}\right).
\end{equation}
Setting $\dot{E}_\mathrm{cool} + \dot{E}_\mathrm{heat} = 0$ allows us to find the steady-state Doppler temperature for sech pulse cooling,
\begin{equation}
    k_\mathrm{B}T_\mathrm{D} = - \frac{\hbar}{\tau_\mathrm{p}} \mathrm{coth}\left( \frac{\delta \, \tau_\mathrm{p}}{2} \right). \label{eq:ClassicalSechDopplerLimit}
\end{equation}

Unlike the lifetime-broadened case (Eq.~(\ref{eq:DopplerLimit})), we find that the Doppler cooling limit with phase-incoherent sech pulses does not exhibit a local minimum in detuning, and is instead minimized for $\delta \rightarrow -\infty$ to $T_\mathrm{min} = \hbar/(\tau_\mathrm{p} k_\mathrm{B}) $.  However, the cooling rate (\ref{eq:SechCoolingRate}) falls off with detuning as $4 \mathrm{e}^{\delta \tau_\mathrm{p}}$, so it may not be practical to detune much further than $\delta = \frac{1}{\tau_\mathrm{p}} \ln (2 - \sqrt{3})$, the detuning that maximizes the damping coefficient and cooling rate.

\section{Quantized molecules}
To see how rotational molasses works on quantized rotors, we first consider the simplest case of a linear molecule in a ${}^1\Sigma$ ground state, for which the electrons don't appreciably perturb the rotational structure.  The cooling will be performed on an electric-dipole-allowed  ${}^1\Sigma \leftrightarrow {}^1\Sigma$ electronic transition that has negligible impact on the geometry (\textit{i.e.}~the rotational constants of the ground and excited states will be assumed to be identical, which is approximately true for the species considered most suitable for translational Doppler cooling, particularly in the large molecule limit).  As mentioned above, we assume that the electronic transition is effectively decoupled from molecular vibration, either through favorable molecular properties or by applying multiple frequencies of the cooling light to repump population from vibrationally excited back to the vibrational ground state.

\begin{figure}
    \centering
    \includegraphics[width=0.45\textwidth]{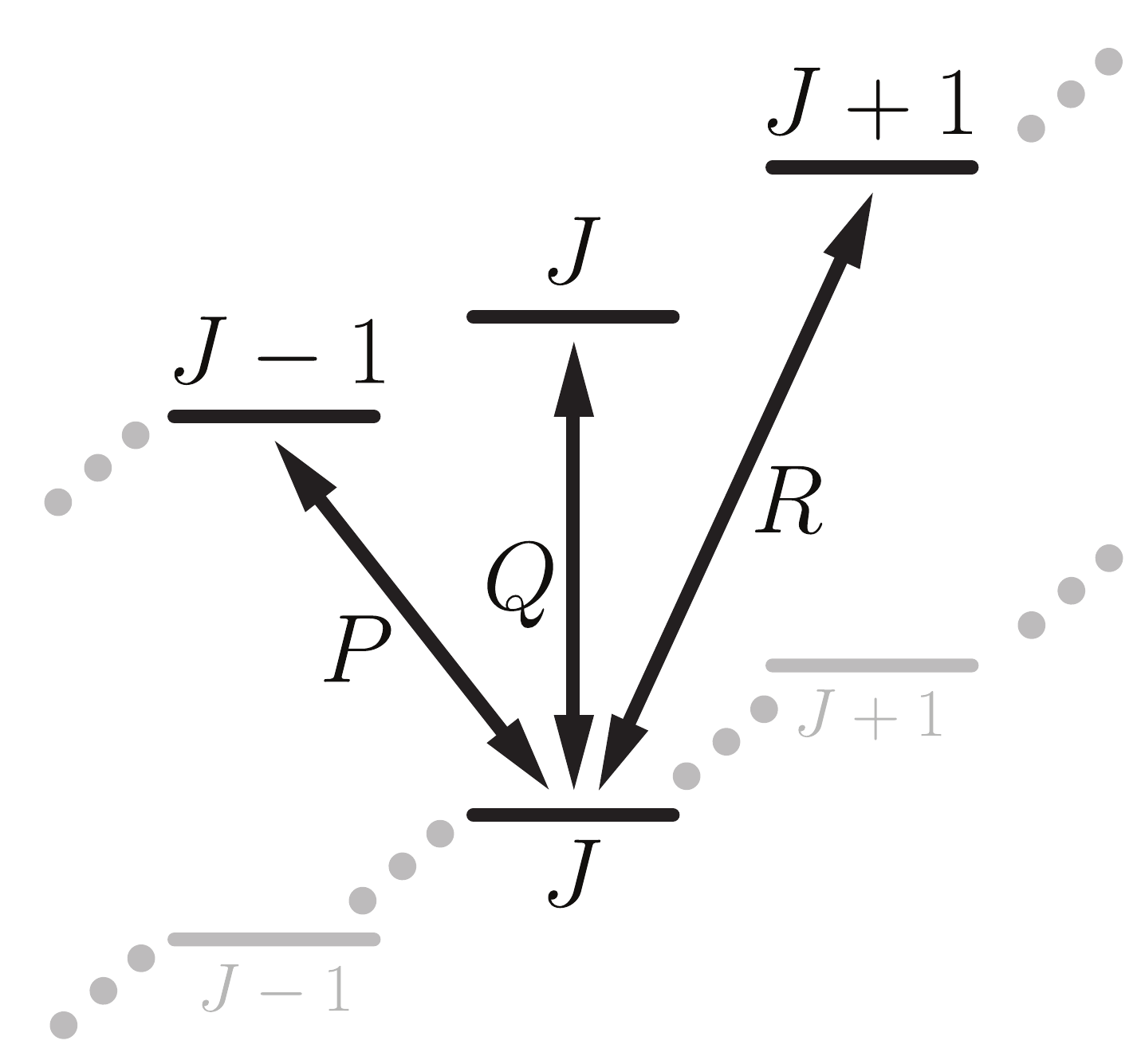}
    \caption{$P$, $Q$, and $R$ branches of an electronic transition in a spherical top or linear rigid rotor molecule.}
    \label{fig:Branches}
\end{figure}

For a linear rotor, the rotational Hamiltonian is $H_\mathrm{rot} = h B J(J+1)$.  Transitions with $\Delta J \equiv J_\mathrm{excited} - J_\mathrm{ground} = \pm 1$ are allowed, with $\Delta J = -1$ known as a ``P branch'' transition and $\Delta J = +1$ known as an ``R branch'' transition (the ``Q branch,'' for which $\Delta J =0$, is forbidden for a $^1\Sigma \leftrightarrow{ {}^1\Sigma}$ system), as shown in Fig.~\ref{fig:Branches}.  The form of $H_\mathrm{rot}$ shows that the P branch transitions will be $4\pi B J$ to the red of the (in this case forbidden) $\Delta J = 0$ band origin, while the R branch transitions will be $4 \pi B(J+1)$ to the blue.  The capture rotational quantum number will therefore be \begin{equation}
 J_\mathrm{c} =\frac{\gamma}{4 \pi B}. \label{eq:CaptureJ}
\end{equation}

We will model the scattering rates on the P and R branches with the steady-state expressions for single two-level systems and single components of the laser polarization ($\hat{\boldgreek{\epsilon}}$) along the quantization direction, with $p=(-1,0,1)$ indexing the $(\sigma^-,\pi,\sigma^+)$ components of the electric field,
\begin{align}
    \Gamma_{\!\mathrm{P}}(J,M,p) =& \,\, \frac{\frac{\gamma}{2}s_\mathrm{P}(J,M,p)}{1 + s_\mathrm{B} + \left( \frac{2 ( \delta + 4 \pi B J )}{\gamma}\right)^2} \label{eq:GammaP}\\
    \Gamma_{\!\mathrm{R}}(J,M,p) =& \,\, \frac{\frac{\gamma}{2}s_\mathrm{R}(J,M,p)}{1 + s_\mathrm{B} + \left( \frac{2 ( \delta - 4 \pi B (J+1) )}{\gamma}\right)^2}.\label{eq:GammaR}
\end{align}
Here, $s_\mathrm{B}$ is an effective saturation parameter that takes into account the power broadening effect of all of the light on the molecule and its couplings between the ground state and both excited states.
We need to consider that the resonant saturation parameter in the numerator, $s_i = \frac{2}{\gamma^2}| \Omega_i |^2$, is a function of the branch type (P and R), the ground quantum state ($J$ and $M$), and the laser polarization component ($p$), which is given by
\begin{align}
    s_\mathrm{P}(J,M,p) = &\,\,\frac{4|\mu_\mathrm{ev}|^2 I_0 (\hat{\boldgreek{\epsilon}}^\ast \!\!\cdot \hat{\mathbf{e}}_{\shortminus p} )^2}{\hbar^2 c \,\varepsilon_0\,\gamma^2}  \nonumber \\
    & \hspace{1cm}\times \left|  \bra{J\!-\!1, M\!+\!p\,} D^{(1)\ast}_{p0}\ket{J,M} \right|^2 \nonumber \\
    = & \,\, \frac{ I_0 }{I_\mathrm{sat}}(\hat{\boldgreek{\epsilon}}^\ast\!\! \cdot \hat{\mathbf{e}}_{\shortminus p} )^2\, J \WignerThreej{J-1}{1}{J}{-M-p}{p}{M}^{\!2} \label{eq:sPLinRot}\\
    s_\mathrm{R}(J,M,p) = & \,\,\frac{4|\mu_\mathrm{ev}|^2 I_0 (\hat{\boldgreek{\epsilon}}^\ast\!\! \cdot \hat{\mathbf{e}}_{\shortminus p} )^2 }{\hbar^2 c \,\varepsilon_0\,\gamma^2} \nonumber \\
    & \hspace{1cm}\times \left|  \bra{J\!+\!1, M\!+\!p\,}  D^{(1)\ast}_{p0} \ket{J,M} \right|^2 \nonumber\\
        =  & \,\, \frac{ I_0  }{I_\mathrm{sat}}(\hat{\boldgreek{\epsilon}}^\ast\!\! \cdot \hat{\mathbf{e}}_{\shortminus p} )^2 (J\!+\!1) \WignerThreej{J+1}{1}{J}{\!-M-p}{p}{M}^{\!2}\!.\label{eq:sRLinRot}
\end{align}
Here, $\gamma \equiv 1/\tau$ is full spontaneous decay width of the electronic excited state (assumed to be independent of $J$) and $I_0$ is the total intensity of the laser light (and therefore $I_0 (\hat{\boldgreek{\epsilon}}^\ast\!\! \cdot \hat{\mathbf{e}}_{\shortminus p} )^2$ is the intensity associated with just the $p$-polarized component).  To get lines (\ref{eq:sPLinRot}) and (\ref{eq:sRLinRot}), we have used the fact that the body-fixed electronic and vibrational transition electric dipole matrix element (which must be along the $q=0$ internuclear axis for a ${}^1\Sigma \leftrightarrow {}^1\Sigma$ transition) is related to the linewidth via
\begin{equation}
    |\mu_\mathrm{ev}|^2 \equiv | \bra{\psi_\mathrm{vib}^\prime} \bra{\psi_\mathrm{el}^\prime} T^{(1)}_{q=0}(\boldgreek{\vec{\mu}}) \ket{\psi_\mathrm{el}}\ket{\psi_\mathrm{vib}} |^2 = \frac{3 \pi \varepsilon_0 \hbar c^3}{\omega^3} \gamma
\end{equation}
and we define the 2-level rotationless saturation intensity as
\begin{equation}
    I_\mathrm{sat} \equiv \frac{\hbar \omega^3}{12 \pi c^2}\gamma
\end{equation}
to write each in terms of $I_\mathrm{sat}$.

Since Eqs.~(\ref{eq:sPLinRot}) and (\ref{eq:sRLinRot}) do not include possible interference between multiple $M$ states being coupled to a single excited state, they implicitly assume that the applied light polarization is made entirely of a single component $p$.  This assumption is valid if the polarization is switched between components in time so that only one is on at any instant, a technique that would potentially be applied anyway to destabilize coherent dark states \cite{Berkeland2002Destabilization}. We will assume that the applied cooling light switches between $p=+1$, $p=0$, and $p=-1$ on the timescale of a couple of scattering events.  Geometrically, this will require at least two beams -- pure $\sigma^{\pm}$ light can be provided by a beam with $\mathbf{k}$ to be parallel to the quantization axis and pure $\pi$ light can be produced by a beam that propagates perpendicular to the first.  The polarization switching 
can be quantitatively accounted for by replacing $(\hat{\boldgreek{\epsilon}}^\ast \cdot \hat{\mathbf{e}}_{-p} )^2 \rightarrow 1$ and multiplying Eqs.~(\ref{eq:GammaP}) and (\ref{eq:GammaR}) by $\frac{1}{3}$.  In this case, the average resonant saturation parameter is given by the sum over polarization components,
\begin{align}
{s_\mathrm{P}} = & \,\, \frac{ I_0 }{I_\mathrm{sat}} \sum_p J \WignerThreej{J-1}{1}{J}{-M-p}{p}{M}^2 = s_0 \frac{J}{2J+1} \label{eq:sP}\\
{s_\mathrm{R}} = & \,\, \frac{ I_0 }{I_\mathrm{sat}} \sum_p (J+1) \WignerThreej{J+1}{1}{J}{-M-p}{p}{M}^2 = s_0 \frac{J+1}{2J+1} \label{eq:sR}
\end{align}
where $s_0 \equiv I_0/I_\mathrm{sat}$.

For spontaneous emission, the analogous angular factor for decay from an excited state with total angular momentum quantum number $J^\prime$ to a ground state with $J=J^\prime +1$ is $(J^\prime+1)/(2 J^\prime +1)$ and the factor for decay to a ground state with $J=J^\prime - 1$ is $J^\prime/(2J^\prime+1)$.

Starting from a ground state with total angular momentum quantum number $J$, absorption on the P(R) branch followed by emission on the R(P) branch will reduce(increase) $J$ by 2.  We can write the power transferred from the laser to the molecules as
\begin{align}
    \dot{E} =& \,\, 2 h B(J+2)\Gamma_{\!\mathrm{R}} - 2hB(J-1) \Gamma_{\!\mathrm{P}} \nonumber \\
    =&\,\, h B \frac{\gamma s_0}{3(2J+1)} \Bigg( \frac{(J+2)(J+1)}{1+ s_\mathrm{B} + \frac{4}{\gamma^2}(\delta - 4 \pi B(J+1))^2} \nonumber \\
    &-   \frac{J(J-1)}{1+ s_\mathrm{B} + \frac{4}{\gamma^2}(\delta + 4 \pi BJ)^2} \Bigg) \label{eq:CoolingPower1}.
\end{align}
We can simplify Eq.~(\ref{eq:CoolingPower1}) assuming $4 \pi B (J+1) \ll |\delta|$,
\begin{align}
    \dot{E} & =  \frac{\shrinkify{\frac{1}{3}} \gamma^3 s_0}{\gamma^2(1 + s_\mathrm{B}) + 4 \delta^2} \nonumber \\
    & \hspace{0.5cm} \times \left( 2hB + hB\left(J(J+1) + 2 \right)\frac{32 \pi \delta B}{\gamma^2(1 + s_\mathrm{B}) + 4 \delta^2} \right),
\end{align}
which allows us to identify the energy damping time constant,
\begin{equation}
    \frac{1}{\tau_E} = -\delta \frac{\frac{4}{3} \hbar \gamma^3 s_0 }{[\gamma^2(1 + s_\mathrm{B}) + 4 \delta^2]^2} \left( \frac{8 \pi B}{\hbar} \right).
\end{equation}
Here, the term in parentheses on the left can be identified as $2/\mu$ from the general treatment in the previous section, and we find that the momentum damping coefficient is smaller than the classical model by a factor of 6: 3$\times$ from the polarization switching and $2\times$ from the rotation of the dipole moment in the lab frame (see Eqs.~(\ref{eq:sP}-\ref{eq:sR})).

We find the equilibrium $\langle J\rangle \equiv J_0$ by setting $\dot{E}=0$, which gives
\begin{align}
    hBJ_0(J_0 + 1) = - \frac{\hbar \gamma}{4} \left( \left( 1 + s_\mathrm{B} \right) \frac{\gamma}{2 \delta} + \frac{2 \delta}{\gamma}  \right) - 2hB \label{eq:FixedPoint1}
\end{align}
where the last term can be neglected in the regime of unresolved rotation ($\gamma \gg 4 \pi B J_0$).  Unlike the isotropic classical model above, this linear molecule only has 2 degrees of freedom, and setting Eq.~(\ref{eq:FixedPoint1}) equal to $k_\mathrm{B}T_\mathrm{D}$ shows that the rotational temperature converges to the same Doppler limit, Eq.~(\ref{eq:DopplerLimit}).

For molecules for which the ground state is still ${}^1\Sigma$ but the excited state is ${}^1\Pi$, the main aspect of this that needs to be adapted is the addition of an allowed Q branch transition ($\Delta J=0$, see Fig.~\ref{fig:Branches}).  Following the same analysis, we find that the Doppler limit is the same as the ${}^1\Sigma \leftrightarrow {}^1\Sigma$ case, Eq.~(\ref{eq:DopplerLimit}).

In fact, this can be generalized to arbitrary singlet electronic states by writing the resonant saturation parameters in terms of the H\"onl-London factors (denoted by script $S$, $\mathscr{S}$) for parity eigenstates ($\epsilon = +1$($-1$) for $e$($f$) levels),
\begin{equation}
    s = s_0\frac{\mathscr{S}_{J'\Lambda'\epsilon';J\Lambda\epsilon}}{(2J+1)}
\end{equation}
where unprimed terms are for the ground state, single primes denote the excited state, and the H\"onl-London factors are given by \cite{Hansson2005Comment}
\begin{align}
    \mathscr{S}_{J'\Lambda'\epsilon';J,\Lambda\epsilon} &\equiv (1 + \delta_{\Lambda'0} + \delta_{\Lambda0} - 2 \delta_{\Lambda'0}\delta_{\Lambda0}) \nonumber \\
    & \hspace{0.4cm} \times(2J'+1)(2J+1) \WignerThreej{J'}{1}{J}{-\Lambda'}{\Lambda'-\Lambda}{\Lambda}^{\!2}\!.
\end{align}

Likewise, the spontaneous emission branching ratios from excited level $J'$,$\epsilon'$ are
\begin{equation}
    \frac{\mathscr{S}_{J'\Lambda'\epsilon';J\Lambda\epsilon}}{\sum_{J'',\epsilon''} \mathscr{S}_{J'\Lambda'\epsilon';J''\Lambda\epsilon''}} = \frac{\mathscr{S}_{J'\Lambda'\epsilon';J\Lambda\epsilon}}{(1+ \delta_{\Lambda'0}\delta_{\Lambda 1})(2J'+1)}.
\end{equation}
By following the same procedure as was explicitly outlined for the ${}^1\Sigma \leftrightarrow {}^1\Sigma$ case, we find that Doppler limit for the arbitrary ${}^1 \Lambda' \leftrightarrow {}^1\Lambda$ case can be written
\begin{equation}
    T_\mathrm{D} = -\frac{\hbar \gamma}{4 k_\mathrm{B}} \left((1 + s_\mathrm{B})\frac{\gamma}{2\delta} + \frac{2\delta}{\gamma}  \right) - \frac{2 hB}{k_\mathrm{B}}w_{\Lambda'\Lambda}
    \label{eq:DopplerLimitGeneral}
\end{equation}
where  $w_{\Lambda'\Lambda}$ is a dimensionless function of $\Lambda$ and $\Lambda'$.  As before, in the regime of unresolved rotation we can typically neglect the second term, and we recover Eq.~(\ref{eq:DopplerLimit}).  Cooling from an initial temperature of $T_0$ to $T_\mathrm{D}$ will increase the peak phase space density by a factor of $T_0/T_\mathrm{D}$.

\section{Numerical Model}

We have conducted numerical simulations of photon spin molasses by solving for the evolution of rotational populations in the presence of cooling laser beams. Energy levels and transition strengths computed by \pgopher~\cite{Western2017} are used to generate transformation matrices for time propagation. The use of \pgopher{} makes it straightforward to rapidly explore a variety of molecular symmetries and parameters.

\subsection{Narrow-band Cooling}

\begin{figure}
    \centering
    \includegraphics[width=1\columnwidth]{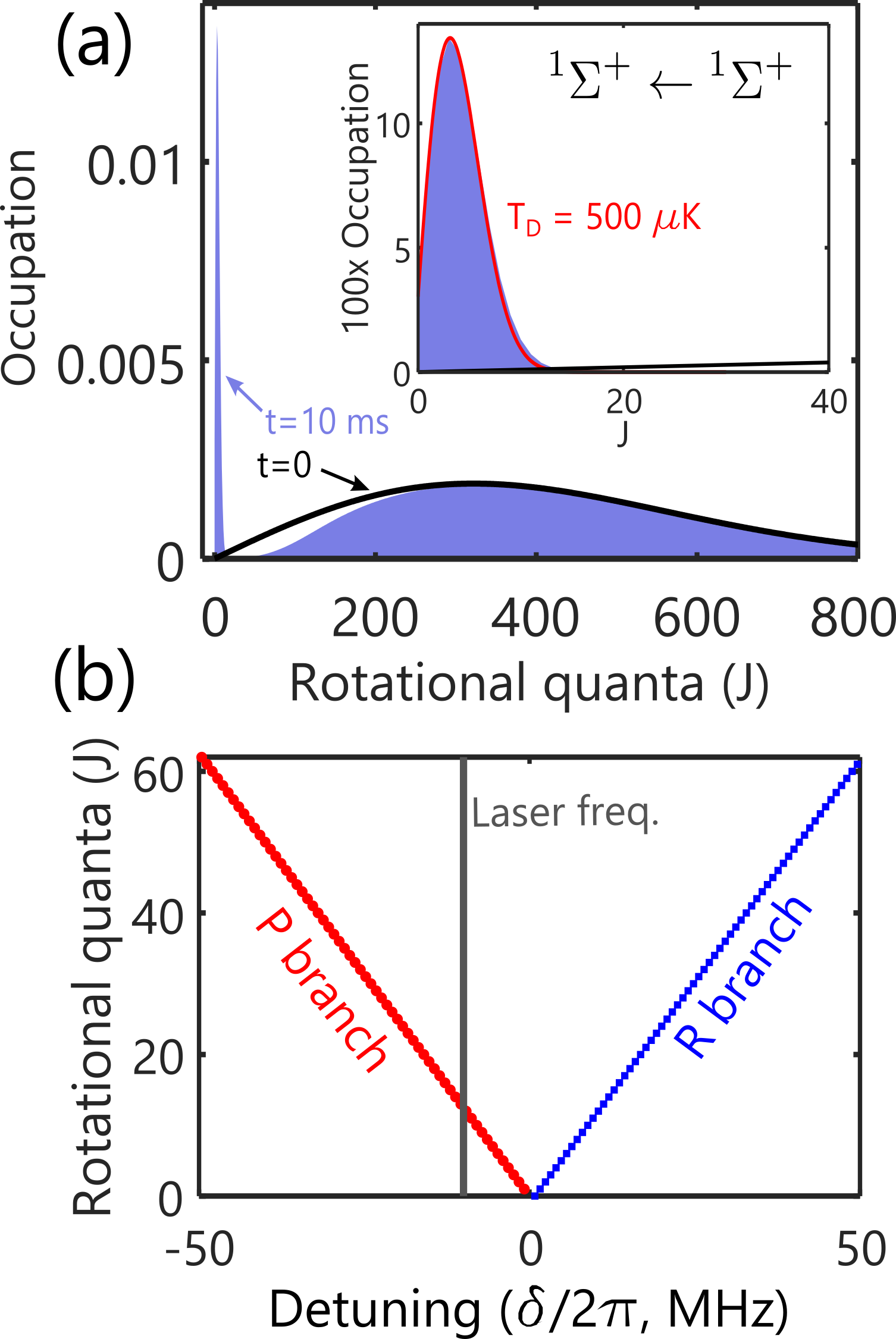}
    \caption{(a) Evolution of rotational distribution for photon spin molasses on a \SigmaSigma{} band. The initial distribution is taken to be at $T=4$~K, as indicated by the black line. The cooled distribution, after 10~ms of photon spin molasses, is shown by the shaded region. Simulation parameters are chosen as $s_0 = 0.1$, $\gamma/2\pi = 20 \mbox{ MHz}$, $\delta = -0.5\gamma$, and $B^{\prime\prime}=B^\prime=0.4$~MHz.  Inset: detail of distributions for $J < 40$. The red line shows a Boltzmann distribution predicted by the Doppler formula Eq.~(\ref{eq:DopplerLimit}). (b) Fortrat diagram corresponding to these cooling conditions. Allowed P and R branch transition frequencies are plotted for each $J$, and the laser frequency is shown relative to the band origin.}
    \label{fig:DistributionsSigmaSigma}
\end{figure}

First consider narrow-band cooling on a \SigmaSigma{} band with an initial rotational distribution at $T_0\approx 4$~K, a realistic value for sources of polyatomic species~\cite{Piskorski2014}. The rotational distribution before and after 10~ms of photon spin molasses is shown in Fig.~\ref{fig:DistributionsSigmaSigma}(a). This simulation assumes $s_0 = 0.1$, $\gamma/2\pi = 20$~MHz, $B'=B''=0.4$~MHz, and $\delta = -\gamma/2$. Approximately 10\% of the rotational distribution is cooled to an effective temperature of $\sim500$~$\upmu$K, as expected from Eq.~(\ref{eq:DopplerLimit}).  Molecules with $J \lesssim 30$ are cooled most efficiently, in line with the prediction $J_\mathrm{c} = 25$ from Eq.~(\ref{eq:CaptureJ}). The Fortrat diagram shown in Fig.~\ref{fig:DistributionsSigmaSigma}(b) can provide intuition for these observations. For the case of a \SigmaSigma{} band with equal rotational constants, only P and R branches exist, and each shows a linear relationship between $J$ and frequency. In this example, the laser is detuned to preferentially drive the P branch near $J \approx 12$, leading the rapid cooling of population near this quantum state.

\begin{figure}
    \centering
    \includegraphics[width=1\columnwidth]{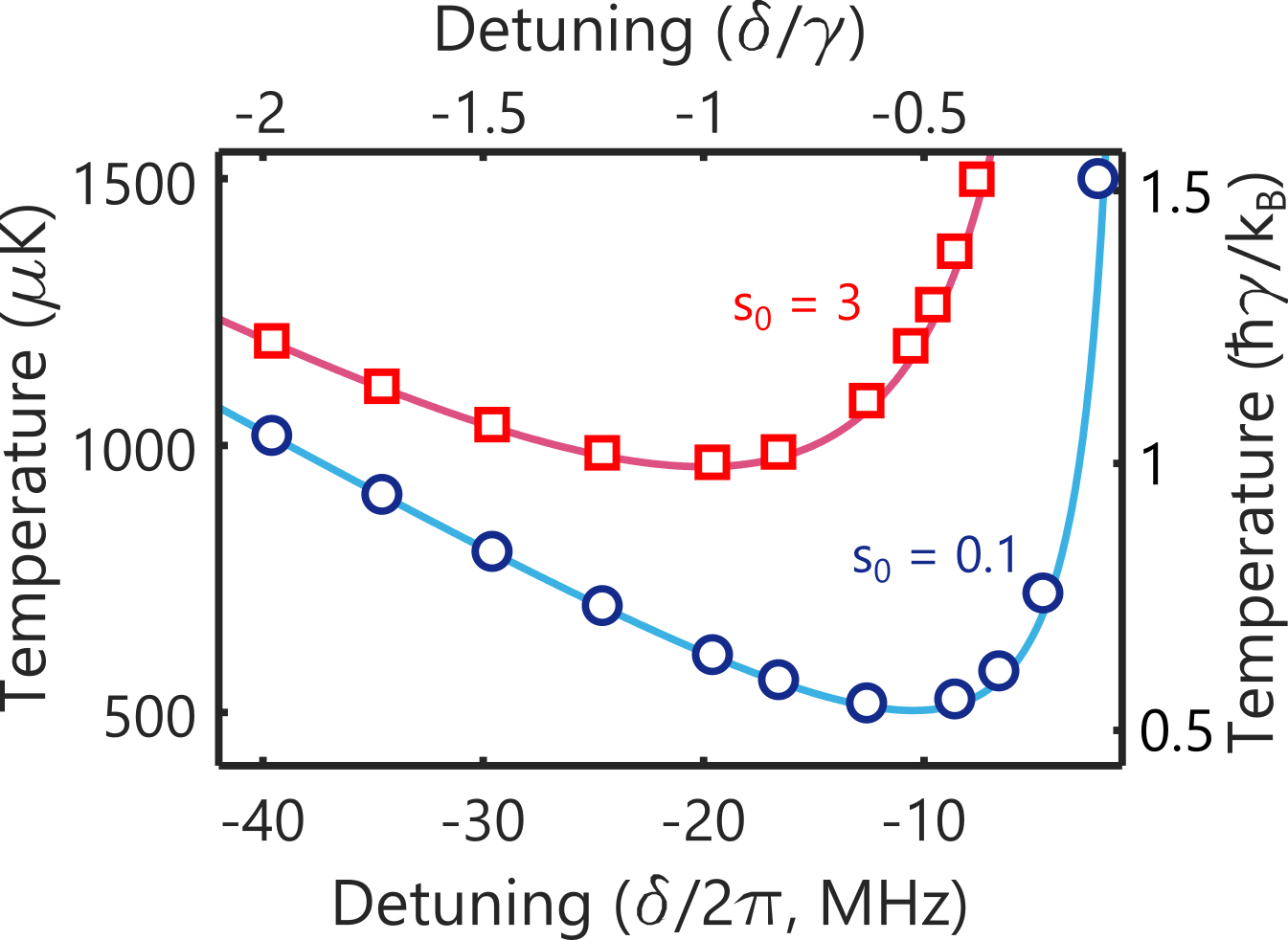}
    \caption{Limiting rotational temperature on a \SigmaSigma{} band as a function of laser detuning for two different intensities. We set $\gamma/2\pi = 20 \mbox{ MHz}$, $B^{\prime\prime}=B^\prime=0.2$~MHz. Blue circles (red squares) correspond to numerical results for $s_0 = 0.1$ ($s_0 = 3$). The corresponding lines are the analytical predictions of Eq.~(\ref{eq:DopplerLimit}).}
    \label{fig:DetuningScan1Sigma}
\end{figure}

The steady-state distribution under these conditions is thermal at $T_\mathrm{f} \approx 500~\upmu$K, although it takes more than 1~s to cool 90\% of the distribution to this value. Figure~\ref{fig:DetuningScan1Sigma} shows the limiting temperature at steady-state under different detuning and intensity. The numerical results agree well with Eq.~(\ref{eq:DopplerLimit}). For the chosen parameters, molecular rotation is cooled to the Doppler temperature typically associated with translational molasses. The results for other combinations of $^1\Sigma$ and $^1\Pi$ ground and excited states show the same limiting temperature, in agreement with analytical predictions. 

\begin{figure}
    \centering
    \includegraphics[width=0.9\columnwidth]{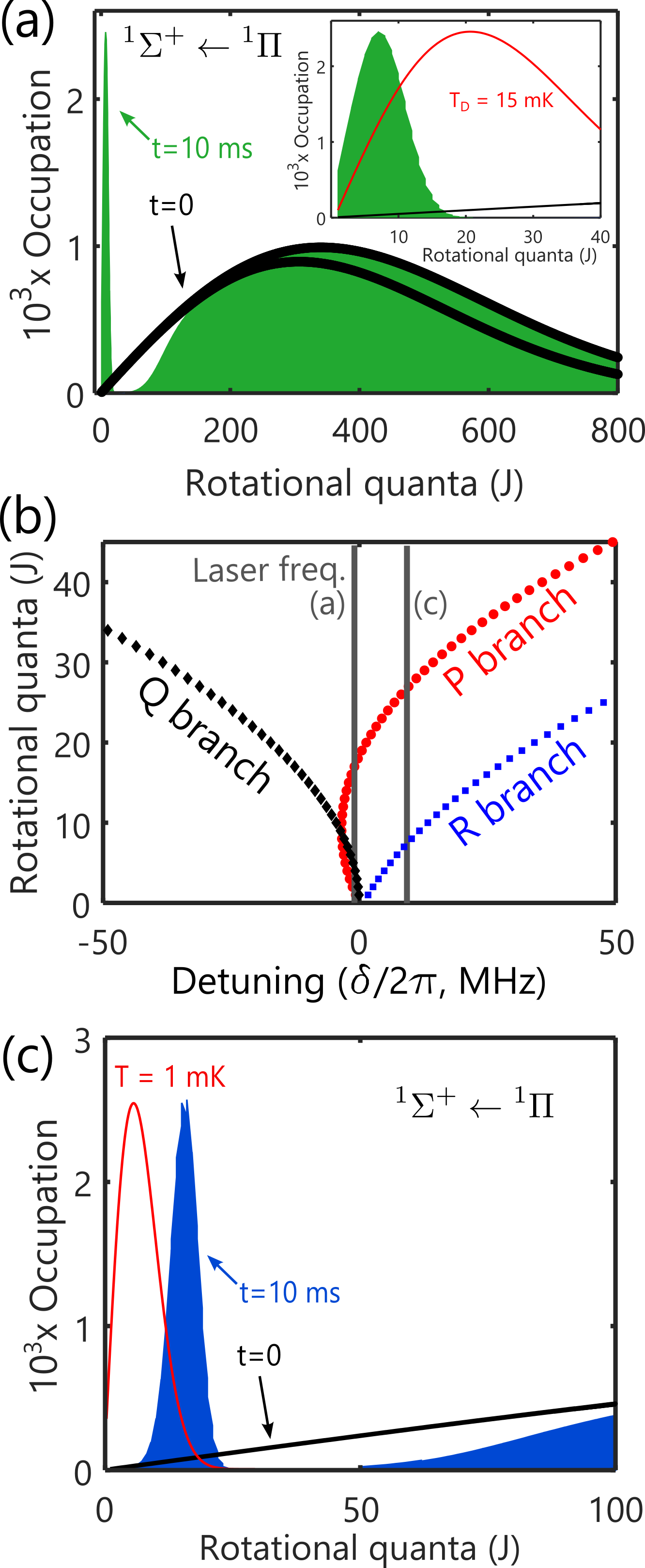}
    \caption{(a) Rotational cooling on a \SigmaPi{} band starting at $T_0=4$~K (black line) and after 10~ms of photon spin molasses (shaded region). Simulation parameters are chosen as $s_0 = 0.3$, $\gamma/2\pi = 20$~MHz, $\delta = -0.01\gamma$, $B''=B'=0.4$~MHz, and $q=-0.0825$~MHz.  Inset: Low-$J$ distribution. Red line shows a Boltzmann distribution predicted by the Doppler formula Eq.~\ref{eq:DopplerLimit}.  (b) Fortrat diagram corresponding to these cooling conditions. (c) Simulated distribution with $\delta = +0.5\gamma$, but other parameters unchanged. To set a scale for the distribution's width, we show a thermal distribution at 1~mK.}
    \label{fig:DistributionsSigmaPi}
\end{figure}

As an example of the features that can appear in more complex level structures, we consider cooling on a \SigmaPi{} transition. In this case, a Q branch exists and the ground state can exhibit parity doubling (characterized by a parameter $q$). If $q=0$ and $\delta=-0.5\gamma$, population is cooled to the same limiting temperature that was found above for \SigmaSigma{} bands ($\sim$500~$\upmu$K). An example with equal rotational constants ($B''=B'=0.4$~MHz) but nonzero parity doubling ($q=-0.0825$~MHz) is shown in Fig.~\ref{fig:DistributionsSigmaPi}(a) for $\delta = -0.01\gamma$. Here, the cooled part of the distribution is characterized by a temperature $T \approx 1$~mK, even though the detuning is set to $\delta = -0.01\gamma$. The prediction of Eq.~(\ref{eq:DopplerLimit}) at $\delta = -0.01\gamma$ is over an order of magnitude larger, $\sim \! 15$~mK. The reason for the lower simulated temperature can be understood from the Fortrat diagram, Fig.~\ref{fig:DistributionsSigmaPi}(b). The frequency dependence of the P and R branches shows that, even when $\delta \approx 0$, the P branch is preferentially driven over the range $J\lesssim 20$. This allows more efficient cooling than predicted by Eq.~(\ref{eq:DopplerLimit}) for such small detunings. Similar reasoning shows that, for these molecular parameters, cooling can be achieved even at blue detuning. For example, when $\delta = +\gamma/2$, the R branch is strongly driven for $J \sim 5-10$ and the P branch is targeted near $J\sim25$. Figure~\ref{fig:DistributionsSigmaPi}(c) shows the cooled distribution under these conditions. Population indeed accumulates around finite $J \sim 16$, but cooling still occurs and the peak phase-space density (PSD) is increased.

\begin{figure}
    \centering
    \includegraphics[width=1\columnwidth]{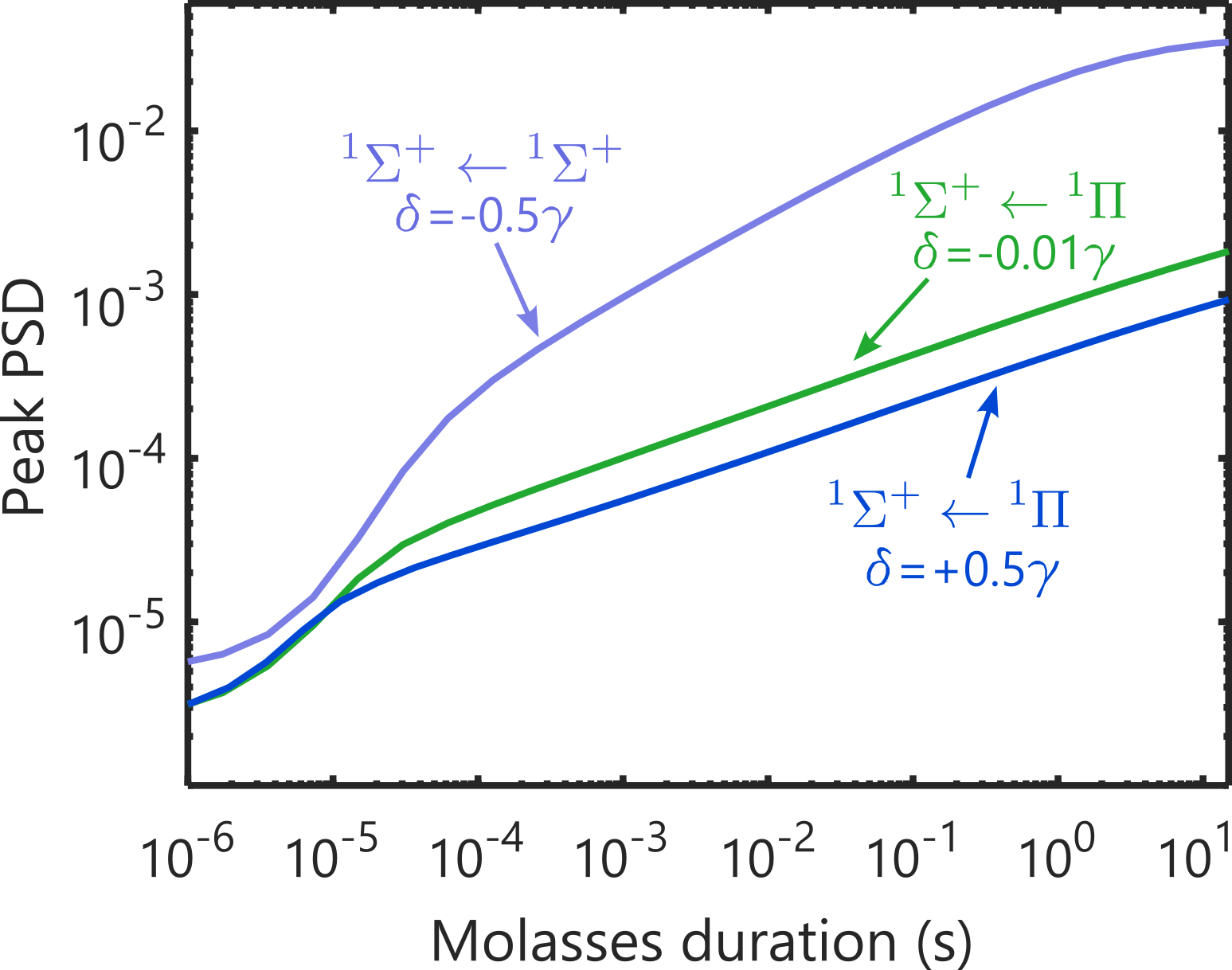}
    \caption{Evolution of peak phase-space density (PSD) as a function of cooling time. Initial conditions and band types are the same as the distributions simulated in Figs.~\ref{fig:DistributionsSigmaSigma} and \ref{fig:DistributionsSigmaPi}, as indicated by the labels.} 
    \label{fig:PeakPSD}
\end{figure}

Phase space compression is a key feature of photon spin molasses. We can characterize this through the peak PSD, computed from the single quantum state with maximum occupation. To demonstrate this compression, Fig.~\ref{fig:PeakPSD} shows the peak PSD as a function of time for the sets of parameters that were used in the simulations described above. Within the first $\sim$50~$\upmu$s after the molasses is turned on, there is a relatively rapid increase in peak PSD as a cooled peak forms centered around $J_0$. This is followed by a slower increase in peak PSD while the rest of the distribution is cooled toward $T_\mathrm{f}$ ($\sim$0.5-1~mK for the parameters chosen). At long times, the simulated rotational distributions approach steady state and the peak PSD saturates. For the \SigmaSigma{} band, this is a thermal distribution at $T_\mathrm{f} \approx 550$~$\upmu$K and the peak PSD is found to be a factor of approximately 7500 larger than the initial peak PSD. This is in line with the expectation of enhancement by a factor of $T_0/T_\mathrm{D} \approx 8000$. For the \SigmaPi{} band exposed to molasses at $\delta = -0.01\gamma$ ($\delta=+0.5\gamma$), the peak PSD ultimately increases by a factor of about 2000 (1000).

\subsection{Broadband Cooling}
To simulate the behavior of cooling via wide-bandwidth light, we simulate applying a series of discrete excitation pulses to a molecule. To avoid inter-pulse coherence effects, we assume that the pulse repetition period, $T_\mathrm{r}$, is kept much longer than the decay time $1/\gamma$~\cite{Ip2018Phonon}. A typical choice is $T_\mathrm{r} = 7/\gamma$ and in our model, no population remains in the excited state at the end of each cycle.

\begin{figure}
    \centering
    \includegraphics[width=1\columnwidth]{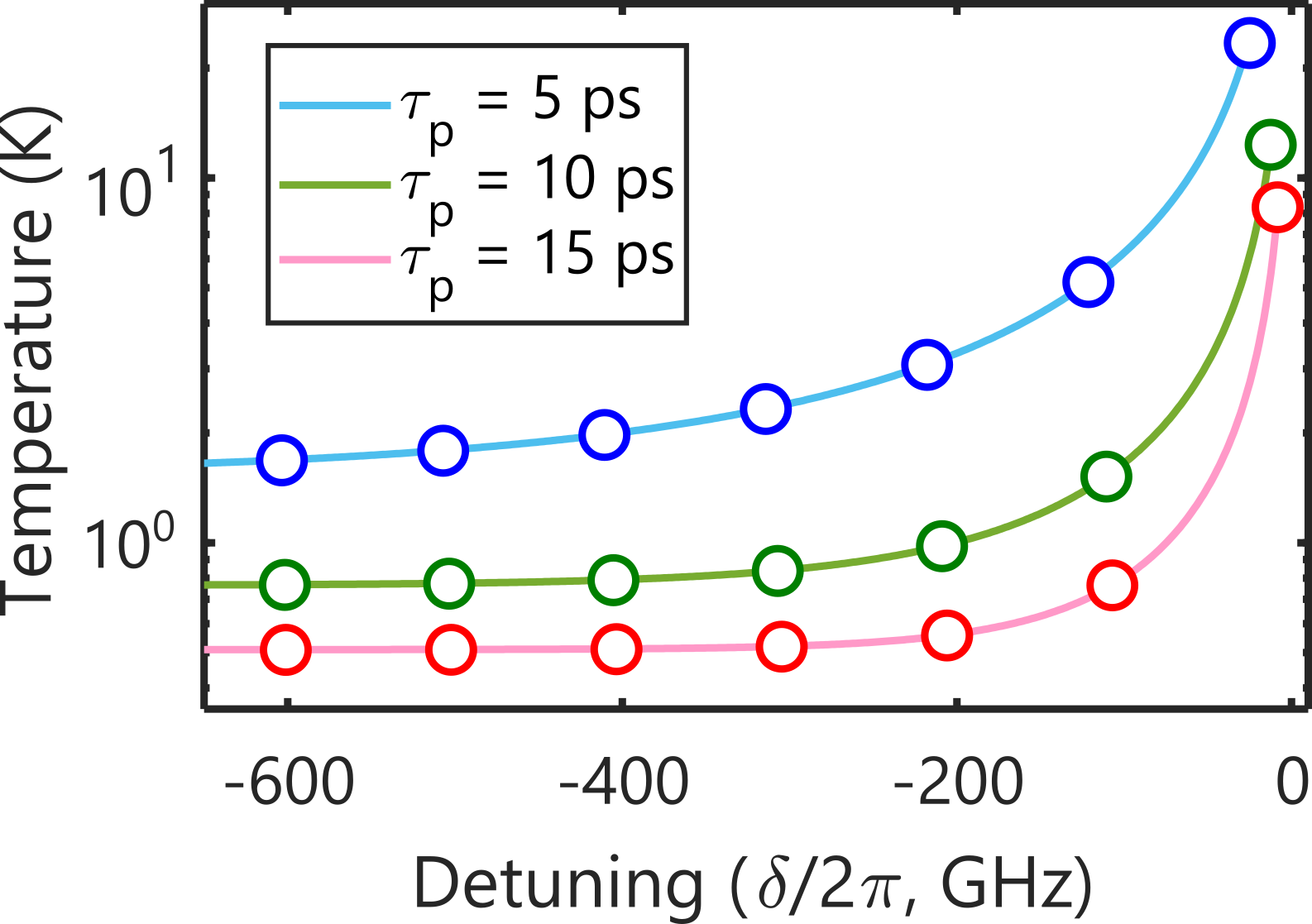}
    \caption{Steady-state temperature for broadband cooling on a \SigmaSigma band and various transform-limited pulse durations, $\tau_\mathrm{p}$. Points are from numerical simulations, solid lines are given by Eq.~(\ref{eq:ClassicalSechDopplerLimit}).}
    \label{fig:SechCoolDopplerScan}
\end{figure}

Figure~\ref{fig:SechCoolDopplerScan} shows the simulated steady-state temperature under different detunings and pulse durations. For concreteness, these simulations assume a \SigmaSigma{} band, rotational constants $B''=B'=20$~MHz and pulse area $\theta_0 = \pi/8$. The limiting temperature decreases monotonically with increasing detuning and agrees with Eq.~(\ref{eq:ClassicalSechDopplerLimit}). For $\tau_\mathrm{p}\approx 10 \mbox{ ps}$ pulses, the Doppler limiting temperature is $T \approx 1 \mbox{ K}$.

This technique could be useful for cooling small molecular ions, including those that appear promising for quantum science applications~\cite{Ivanov2020InSearch, Mills2020Dipole}. At least some of the identified molecules possess features desirable for optical cycling, e.g. short radiative lifetimes and nearly-diagonal Franck-Condon factors (FCFs)~\cite{Ivanov2020InSearch}. Moreover, it is relatively straightforward to trap molecular ions with long trapping lifetimes. As an example, we can consider the specific case of \YFplus. This molecule possess a moderate rotational constant ($B \approx 8.7$~GHz), a strong $C\,^2\Pi \leftarrow X\,^2\Sigma^+$ band ($\gamma/2 \pi \approx  37$~MHz) with predicted diagonal FCF $>0.999$~\cite{Ivanov2020InSearch}, and reasonable transition wavelengths for pulsed-laser excitation ($\Delta E_{C-X} \approx 4.0$~eV). To asses the rotational cooling effect of photon spin molasses in isolation, we neglect the intermediate electronic states that provide alternative decay paths in this species, though these are present in some species that have been successfully laser cooled \cite{Yeo2015Rotational}.

\begin{figure}
    \centering
    \includegraphics[width=1\columnwidth]{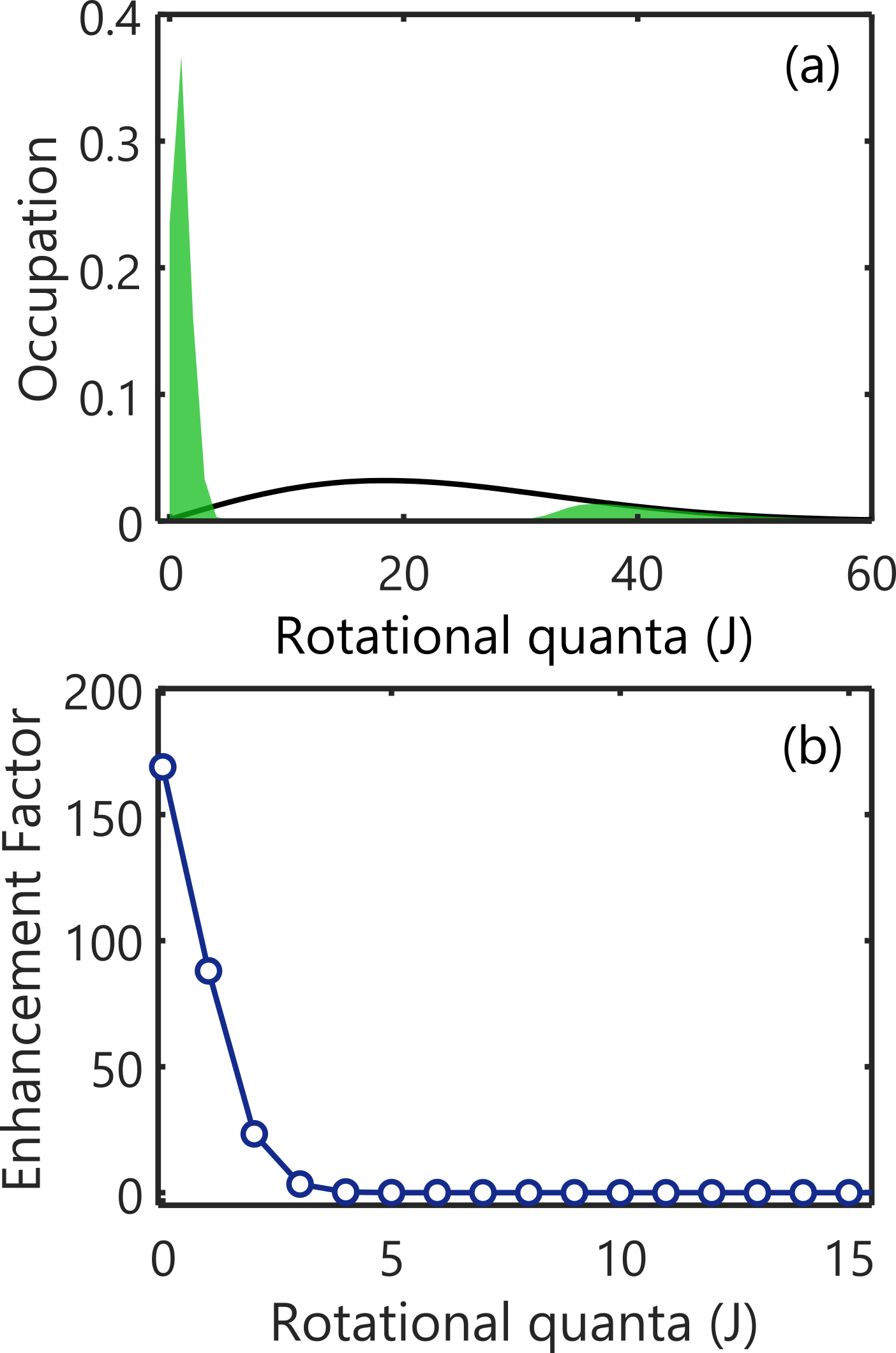}
    \caption{Simulated sech-pulse cooling of \YFplus molecules. (a) Rotational distribution before (black line) and after (shaded region) 20~ms of pulsed cooling. (b) Fractional enhancement of each rotational state relative to the initial distribution.}
    \label{fig:YFPlusCooling}
\end{figure}

In Fig.~\ref{fig:YFPlusCooling}(a) we plot the evolution of a distribution of \YFplus molecules exposed to sech-pulse photon spin molasses. We consider an initial thermal distribtuion at 300~K. The final distribution is shown after 20~ms of photon spin molasses ($\sim 7 \times 10^5$ laser pulses) with $\tau_\mathrm{p} = 6$~ps and $\delta \approx -10/\tau_\mathrm{p}$. The resulting Doppler limit is around 1~K. These laser properties balance frequency bandwidth against Doppler-limited final temperature in order to optimize the population cooled to the lowest few $J$ states. As shown in Fig.~\ref{fig:YFPlusCooling}(b) , the population in the absolute ground state is enhanced by over two orders of magnitude by broadband photon spin molasses.

\section{Discussion}
The photon spin molasses described here is capable of cooling molecular rotation, but other methods have also been proposed and/or demonstrated that may be advantageous depending upon circumstance.  For instance, optical pumping with narrowband \cite{Staanum2010Rotational,Schneider2010AllOptical} or spectrally filtered braodband pulses 
\cite{Lien2014Broadband,Cournol2018Rovibrational,Stollenwerk2020Cooling} has been shown to be effective for cooling molecular rotation in the regime where the spectral branches are rotationally resolved with respect to the natural linewidth.  Photon spin molasses can be thought of as the limiting case of pushing that technique to the unresolved regime. This regime arises as the ratio of the rotational constant to the natural linewidth decreases due to increasing molecular mass and size.  Additionally, since photon spin molasses is not efficient for cooling molecules with rotational quanta above the capture range $J_\mathrm{c}$, it is likely that pre-cooling with another method such as supersonic expansion \cite{Kantrowitz1951High,Abeysekera2014ShortPulse,Aggarwal2021Supersonic}, buffer gas cooling \cite{Patterson2009Intense,Hutzler2012Buffer,Hansen2014Efficient,Skoff2011Diffusion,Gantner2020BufferGas}, or sympathetic collisional cooling by laser-cooled atoms \cite{Hudson2016Sympathetic} would be advantageous and could extend the reach of these techniques to the sub-mK regime.

\textit{Acknowledgments }The authors acknowledge helpful discussions with John Doyle, David Patterson, Paul Hamilton, Eric Hudson, and Amar Vutha. This work was supported by the AFOSR under Grant No.~FA9550-20-1-0323 and the CIQC
Quantum Leap Challenge Institute through NSF award OMA-2016245 and NSF under Grant No.~PHY-1912555. BLA acknowledges financial support from the Gordon and Betty Moore Foundation under award \#7947.

\bibliography{RotationalMolasses}

\end{document}